\documentclass[pageno,nohyperref]{jpaper}

\usepackage{fancyhdr}
\usepackage[normalem]{ulem}
\usepackage[sort,nocompress]{cite}
\usepackage{flushend}
\usepackage{enumitem}

\usepackage{graphics}
\usepackage{xspace}
\usepackage{booktabs}
\usepackage{multirow}
\usepackage{amssymb}
\title{MOD: Minimally Ordered Durable Datastructures for Persistent Memory}

\begin{document}
\newcommand{\itempar}[1]{ \noindent \textbf{#1}}
\newcommand{\mypar}[1]{\vspace{0.02cm} \noindent \textbf{#1}}
\newcommand{\myitem}{\vspace{-0.1cm} \item}
\newcommand{\todo}[1]{\textbf{TODO: [#1]}}
\newcommand{\fixme}[1]{\textbf{FIXME: [#1]}}
\newcommand{\Fixme}{{\bf FIXME:}}
\newcommand{\remark}[1]{\textbf{REMARK [#1]???}} 
\newcommand{\response}[1]{\textbf{RESPONSE [#1]???}} 
\newcommand{\dropme}[1]{\textbf{DROPME [#1]???}} 
\newcommand{\reviewcomment}[2]{\textbf{REVIEW \##1}{\xspace\emph{#2}}} 
\newcommand{\eg}{\textit{e.g.}}
\newcommand{\ie}{\textit{i.e.}}
\newcommand{\etal}{\textit{et al.}\xspace}
\newcommand{\codesm}[1]{\texttt{\small #1}}
\newcommand{\code}[1]{\texttt{#1}}
\newcommand{\ignore}[1]{}
\newcommand{\myincludegraphics}[2]{\resizebox{#1}{!}{\includegraphics{#2}}}
\newcommand{\SmallUpperCase}[1]{\scalebox{0.8}[0.8]{#1}}    

% convenient macros for use with algorithmic
\newcommand*\BitAnd{\mathrel{\&}}
\newcommand*\BitOr{\mathrel{|}}
\newcommand*\ShiftLeft{\ll}
\newcommand*\ShiftRight{\gg}
\newcommand*\BitNeg{\ensuremath{\mathord{\sim}}}

% convenient macros for 
\newcommand{\ofence}{\texttt{ofence}\xspace}
\newcommand{\dfence}{\texttt{dfence}\xspace}
\newcommand{\clwb}{\texttt{clwb}\xspace}
\newcommand{\clwbs}{\texttt{clwbs}\xspace}
\newcommand{\clflush}{\texttt{clflush}\xspace}
\newcommand{\clflushopt}{\texttt{clflushopt}\xspace}
\newcommand{\pcommit}{\texttt{pcommit}\xspace}
\newcommand{\epoch}{\texttt{epoch}\xspace}
\newcommand{\fence}{\texttt{fence}\xspace}
\newcommand{\mov}{\texttt{mov}\xspace}
\newcommand{\movnti}{\texttt{movnti}\xspace}
\newcommand{\sfence}{\texttt{sfence}\xspace}
\newcommand{\mfence}{\texttt{mfence}\xspace}
\newcommand{\lazypcommit}{\texttt{lazy\_pcommit}\xspace}
\newcommand{\pmdk}{\SmallUpperCase{PMDK}}
\newcommand{\lib}{MOD}

\newcommand{\fsfull}{Functional Shadowing\xspace}
\newcommand{\fs}{FS\xspace}

\author{Swapnil Haria, Mark D. Hill, Michael M. Swift\\
           University of Wisconsin-Madison \\
           \{swapnilh,markhill,swift\}@cs.wisc.edu}
\date{}
\maketitle

\thispagestyle{empty}

\begin{abstract}
Persistent Memory (PM) makes possible recoverable applications that can preserve application progress across system reboots and power failures. 
Actual recoverability requires careful ordering of cacheline flushes, currently done in two extreme ways.
On one hand, expert programmers have reasoned deeply about consistency and durability to create applications centered on a single custom-crafted durable datastructure.
    On the other hand, less-expert programmers have used software transaction memory (STM) to make atomic one or more updates, albeit at a significant performance cost due largely to ordered log updates.

In this work, we propose the middle ground of composable persistent datastructures called Minimally Ordered Durable (MOD) datastructures.
MOD is a C++ library of several datastructures---currently, map, set, stack, queue and vector---that often perform better than STM and yet are relatively easy to use. 
They allow multiple updates to one or more datastructures to be atomic with respect to failure. 
Moreover, we provide a recipe to create more recoverable datastructures.

MOD is motivated by our analysis of real Intel Optane PM hardware showing that allowing unordered, overlapping flushes significantly improves performance.
MOD reduces ordering by adapting existing techniques for out-of-place updates (like shadow paging) with space-reducing structural sharing (from functional programming).
MOD exposes a Basic interface for single updates and a Composition interface for atomically performing multiple updates.
Relative to the state-of-the-art Intel PMDK v1.5 STM, MOD improves map, set, stack, queue  microbenchmark performance by 40\%,  and speeds up application benchmark performance by 38\%.

\end{abstract}

\section{Introduction}

Persistent Memory (PM) is here---Intel Optane DC Persistent Memory Modules (DCPMM) began shipping in 2019~\cite{intel-news}.
Such systems expose fast, byte-addressable, non-volatile memory (NVM) devices as main memory and allow applications to access this persistent memory via regular load/store instructions.
In fact, we ran all experiments in this paper on a system with engineering samples of Optane DCPMM~\cite{optane, swanson-preprint}.

\ignore{
\begin{figure}[t!]
    \centering
    \myincludegraphics{3.5in}{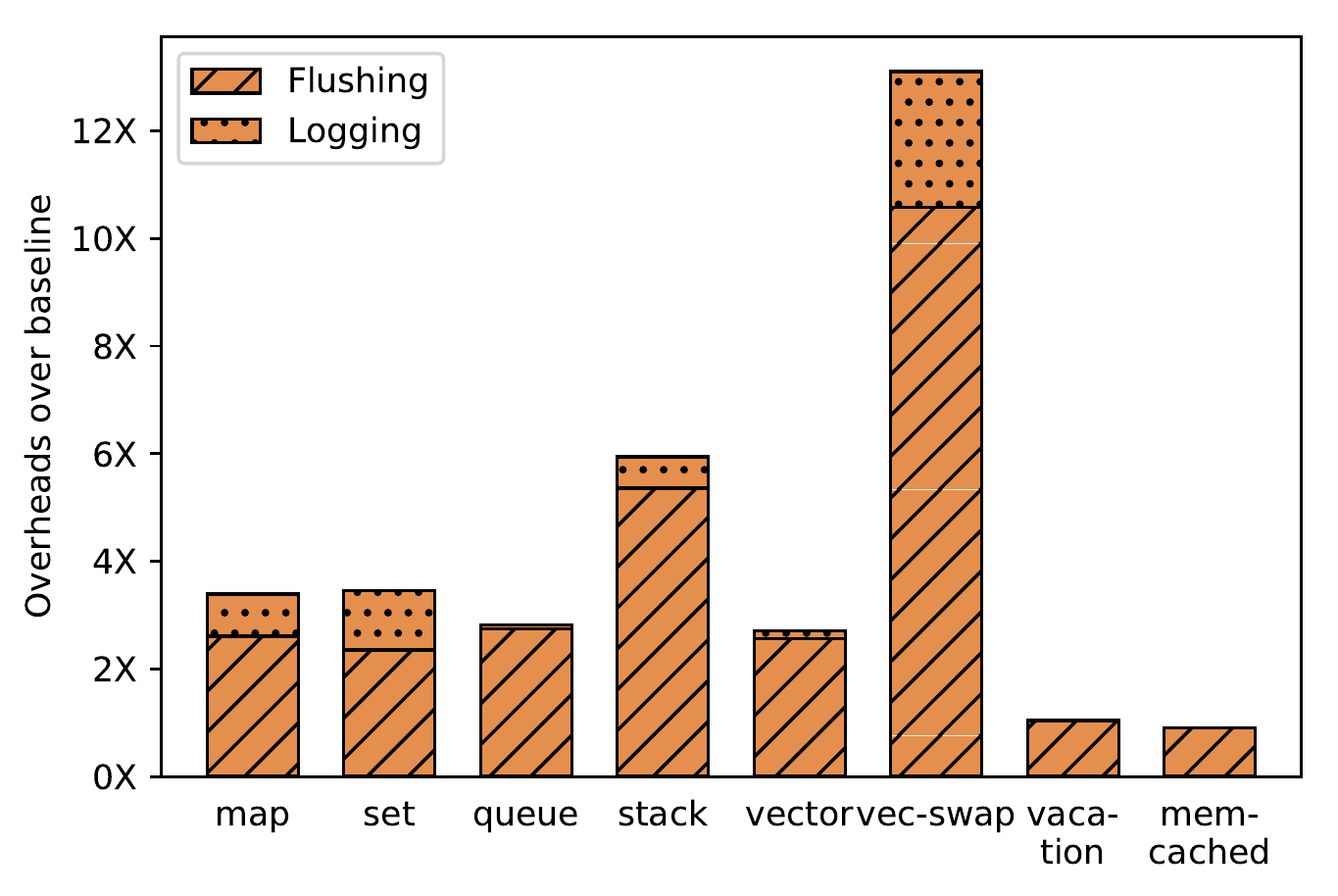}
    \caption{PM-STM overheads in recoverable PM workloads implemented using PMDK v1.5.}
    \label{fig:tx-overheads}
\end{figure}
}

The durability of PM enables \emph{recoverable applications} that preserve in-memory data beyond process lifetimes and system crashes, a desirable quality for workloads like databases, key-value stores and long-running scientific computations~\cite{Caulfield:2010, Li:2012}.
Such applications use cacheline flush instructions to move data from volatile caches to durable PM and order these flushes carefully to ensure consistency.
For instance, applications must durably update data before updating a persistent pointer that points to the data, or atomically do both.

However, few recoverable PM applications have been developed so far, even though PM libraries like Mnemosyne~\cite{Volos:2011} and Intel Persistent Memory Development Kit (PMDK)~\cite{pmdk} have existed for several years.
Currently, there are two approaches to building such applications: single-purpose custom datastructures (e.g., persistent B-trees~\cite{Chen:2015, Venkataraman:2011, Yang:2015}) or general-purpose transactions.
While both approaches have some benefits, we believe that neither is suitable for encouraging developers to start building PM applications.

Although custom datastructures are typically very fast, significant effort is needed in designing these structures to ensure that updates are performed atomically with respect to failure i.e., either all modified data is made durable in PM or none.
Accordingly, the designers need to ensure that modified data is logged, dirty cachelines are explicitly flushed to PM in a deliberate order enforced by the use of appropriate fence instructions for consistency.
Furthermore, performance optimizations useful in one datastructure may not generalize to other datastructures.
These custom datastructures typically do not support the composition of failure-atomic operations spanning multiple datastructures, e.g., popping an element of a durable queue and inserting it into a durable map.

Existing PM libraries offer software transactional memory (PM-STM) for building general-purpose crash-consistent code, but with complicated interfaces and high performance overheads.
Operations on existing datastructures can be wrapped in transactions to facilitate consistent recovery on a crash.
These transactions also allow developers to compose failure-atomic operations that update multiple datastructures.
However, it is not easy to use these PM-STM interfaces correctly.
For instance, the state-of-the-art PMDK transactions require programmer annotations (\code{TX\_ADD}) in each transaction to demarcate memory that could be modified in that transaction.
Incorrect usage of such annotations is a common source of crash-consistency bugs in applications built with PMDK~\cite{pmtest}.

Moreover, the generality of transactions come at a high performance cost.
As we will show, about 64\% of the overall execution time in PM-STM based applications is spent in flushing activity.
These high overheads arise from excessive ordering constraints in these transactions, with each transaction having 5-11 ordering points (i.e., \sfence on x86-64).
Our experiments on Optane DCPMM show that flushes (i.e., \clwb on x86-64) slow execution more when they are more frequently ordered. 
For instance, 8 \clwbs can be performed 75\% faster when they are ordered jointly by a single \sfence than when each \clwb is individually ordered by an \sfence.
 
To make PM application development more widely accessible, we propose a middle ground: Minimally Ordered Durable Datastructures (\lib), a library of many persistent datastructures with simple abstractions and good performance (and a methodology to make more).
\lib~performs better than transactions in most cases and also allows the composition of updates to multiple persistent datastructures. 
To allow the programmer to easily build new PM applications, we encapsulate away the details of persistence such as crash-consistency, ordering and durability mechanisms in the implementation of these datastructures.
Instead, \lib~enables programmers to focus on core logic of their applications.

Similar efforts such as the Standard Template Library (STL)~\cite{stl} in C++ have proved extremely popular, allowing programmers to develop high-performance applications using simple datastructure abstractions whose efficient and complicated implementations are hidden from the programmer.
\lib~offers datastructure abstractions similar to those in the STL, namely map, set, stack, queue and vector.
For each datastructure, \lib~offers convenient failure-atomic update and lookup operations with familiar STL-like interfaces such as \code{push\_back} for vectors and \code{insert} for maps.

New abstractions get wider adoption only if they perform well.
For high performance, \lib~datastructures use shadow paging~\cite{Lorie:1977, Gray:1981} to minimize internal ordering in update operations---one \sfence per failure-atomic operation in the common case.
Specifically, we rely on out-of-place writes to create a new and updated copy (\emph{shadow}) of each datastructure without overwriting the original data.
These out-of-place writes do not need to be logged and can be flushed with overlapping flushes to minimize flushing overheads.

To reduce the memory overhead introduced by shadow paging, our datastructures use \emph{structural sharing} optimizations found in purely functional datastructures~\cite{Driscoll:1989,Okasaki:1998,Puente:2017, Steindorfer, Stucki:2015}.
With these optimizations, the updated shadow is built out of the unmodified data of the original datastructure plus modest new and updated state.
Consequently, the shadow incurs additional space overheads of less than 0.01\% over the original datastructure.
On Intel Optane DCPMM, our \lib~datastructures improve the performance of map, set, stack, queue microbenchmarks by 43\%, hurts vector by 122\%, and speeds up application benchmarks by 36\% as compared to state-of-the-art Intel PMDK v1.5.
We also present a methodology to repurpose other existing purely functional datastructures into new persistent datastructures. 

Finally, \lib~also offers the ability to compose failure-atomic updates to multiple durable datastructures.
Only for such use cases, we expose the underlying out-of-place update operations to the programmer.
Thus, programmers can update multiple datastructures to generate new versions of these datastructures.
We provide a convenient \code{Commit} interface to failure-atomically replace all the original datastructures with their respective updated versions.

We make the following contributions in this paper:
\begin{itemize}[topsep=-1pt]
    \item We develop the design and implementation of the \lib~library of  high-performance durable datastructures with encapsulated persistence.
    \item We present two alternative interfaces for using \lib~datastructures for different use cases.
    \item We provide a recipe to create more \lib~datastructures from existing functional datastructures. 
    \item We develop an analytical model for estimating the latency of concurrent cacheline flushes with Optane DCPMM.
    \item We release a C++ implementation of \lib~datastructures.
\end{itemize}

\begin{figure*}[th]
    \centering
    \myincludegraphics{6in}{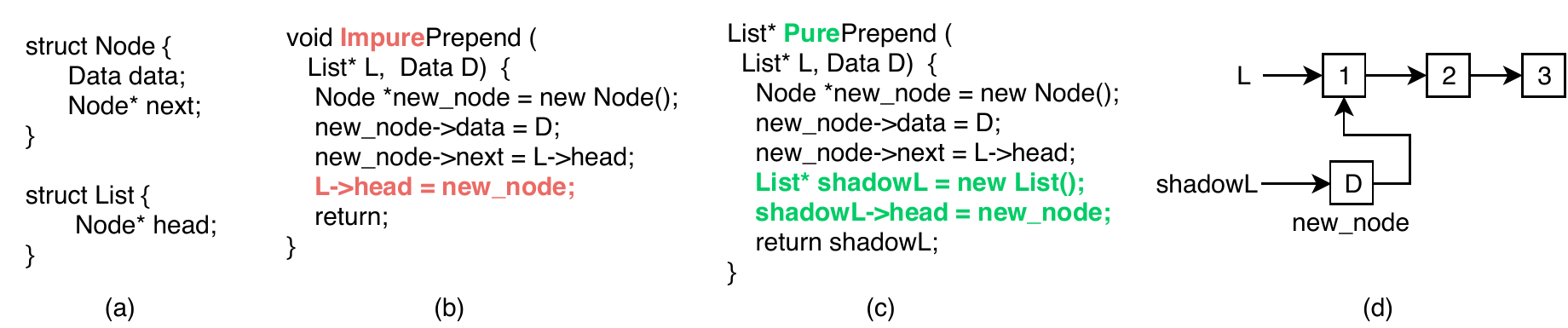}
    %\vspace{-8mm}
    \caption{For linked list defined in (a), implementation of prepend as (b) impure function with original list \code{L} modified and (c) pure function where new updated \code{shadowL} is created and returned. (d) \code{shadowL} reuses nodes of list \code{L} to reduce space overheads.}
    \label{fig:fp-bg}
    %\vspace{-6mm}
\end{figure*}

\section{Background}
\label{background}
We first provide basic knowledge of PM programming and functional programming as required for this paper.

\subsection{Persistent Memory System}
We consider a system in which the physical address space is partitioned into volatile DRAM and durable PM.
While the contents of PM are preserved in case of a system failure, DRAM and other structures such as CPU registers, caches, etc., are wiped clean.
This system model is similar to most prior work~\cite{Chakrabarti:2014, ido, nalli:2017, Volos:2011} and representative of Optane DCPMM.

Recoverable PM software rely on hardware guarantees to know when PM writes are \emph{persisted}, i.e., when a write is guaranteed to be durable in PM.
Writes are first stored in volatile caches to exploit temporal locality of accesses and written back to PM at a later time unknown to software, depending on the cache replacement policy.
Hence, PM systems support two instructions for durability and/or ordering: a flush instruction to explicitly writeback a cacheline from the volatile caches to PM, and a fence instruction to order subsequent instructions after preceeding flushes become durable.
%To guarantee that a PM write is persisted, PM applications issue cacheline flush instructions (e.g., \clflushopt/\clwb in x86-64) to force a writeback of cached data (corresponding to a particular memory address) to PM.
%These cacheline flush instructions 
%After performing one or more cacheline flushes, applications issue persistent fence instructions(e.g., \sfence) that stall the core till all outstanding flushes 

%Naively, writes can only be considered persisted once the store updates PM.
%Certain optimizations such as Asynchronous DRAM Refresh (ADR) allow writes to be considered persisted once they reach the memory controller~\cite{intel-adr}.

%Our Intel test machine supports the \clwb/\clflushopt/\clflush instructions to flush a specific address and the \sfence instruction that stalls till the completion of all \clflushopt/\clwb instructions from the same thread.

\subsection{Persistent Memory Programming}
Here, we discuss applications that rely on the persistence of PM.
Such applications are \emph{recoverable} if they store enough state in PM to successfully recover to a recent state and without losing all progress after a system crash.
There are several challenges involved in programming recoverable applications. 
Sufficient application data must be persisted in PM to allow successful recovery to a consistent and recent state.
System crashes at inopportune moments could result in partially updated and thus inconsistent datastructures that cannot be used for recovery.
As a result, programmers have to carefully reason about the ordering and durability of PM updates.
Unfortunately, PM updates in program order can be reordered by hardware including write-back caches and memory controller (MC) scheduling policies.

To abstract away these programming challenges, researchers have developed \textit{failure-atomic sections} (FASEs)~\cite{Chakrabarti:2014}.
FASEs are code segments with the guarantee that all PM writes within a FASE happen atomically with respect to system failure.
For example, prepending to a linked list (Figure~\ref{fig:fp-bg}b) in a FASE guarantees that either the linked list is successfully updated with its head pointing to the durable new node or that the original linked list can be reconstructed after a crash.

PM libraries~\cite{Coburn:2011, pmdk, Volos:2011} typically implement FASEs with software transactions that guarantee failure-atomicity and durability.
All updates made within a transaction are durable when the transaction commits.
If a transaction gets interrupted due to a crash, write-ahead logging techniques are typically used to allow recovery code to clean up partial updates and return persistent data to a consistent state.
Hence, recoverable applications can be written by allocating datastructures in PM and only updating them within PM transactions.
We discuss the performance bottlenecks of PM transactions in Section~\ref{motivation}.

\subsection{Functional Programming Concepts}
\label{sec:bg-fp}
In this work, we leverage two basic concepts in functional programming languages: pure functions and purely functional datastructures.
These ideas are briefly described below and illustrated in Figure~\ref{fig:fp-bg}.

\mypar{Pure Functions.} 
A pure function is one whose outputs are determined solely based on the input arguments and are returned explicitly.
Pure functions have no externally visible effects (i.e., side effects) such as updates to any non-local variables or I/O activity.
Hence, only data that is newly allocated within the pure function can be updated. 
%Finally, given the same inputs, pure functions always return the same outputs.
Figure~\ref{fig:fp-bg} shows how a pure and an impure function differ in performing a prepend operation to a list.
The impure function overwrites the head pointer in the original list \code{L}, which is a non-local variable and thus results in a side effect.
In contrast, the pure function allocates a new list \code{shadowL} to mimic the effect of the prepend operation on the original list and explicitly returns the new list.
Note that the pure function does not copy the original list to create the new list.
Instead, it reuses the nodes of the original list without modifying them.
%Figure 1 shows an example of an impure function and a pure function which perform the same task.

\mypar{Functional Datastructures.} 
Commonly used in functional languages, purely functional or \emph{persistent datastructures} are those that preserve previous versions of themselves when modified~\cite{Driscoll:1989}.
We refer to these as purely functional datastructures in this paper to avoid confusion with persistent (i.e., durable) datastructures for PM.

Purely functional datastructures are never modified in-place.
Instead, every update of such datastructures creates a logically new version while preserving the old version.
Thus these datastructures are inherently multi-versioned.

To reduce space overheads and improve performance, functional datastructures (even arrays and vectors) are often implemented as trees~\cite{Okasaki:1998,Puente:2017}.
Tree-based implementations allow different versions of a datastructure to appear logically different while sharing most of the internal nodes of the tree.
For example, Figure~\ref{fig:fp-bg} shows a simple example where the original list \code{L} and the updated list \code{shadowL} share nodes labeled 1, 2 and 3.
Such optimizations are called \emph{structural sharing}.

\section{Understanding Performance Bottlenecks}
\label{motivation}
Good performance typically aids the adoption of new abstractions.
Thus in this section, we try to identify the main performance bottlenecks in PM-STM workloads and understand how to mitigate these overheads.

\mypar{Overheads in PM-STM Workloads.}
At a high level, PM-STM implementations suffer from two main overheads: flushing (required for durability of data) and logging (required for failure-atomicity). 
We measured these overheads on Optane DCPMMs by running recoverable PM workloads (described in Table~\ref{benchmarks} in Section~\ref{sec:eval}) with Intel PMDK v1.5, a state-of-the-art PM-STM implementation that uses hybrid undo-redo logging.
As shown in Figure~\ref{fig:tx-overheads}, these applications spend on average about 64\% of their execution time performing flushing and 9\% performing logging operations.
These PM-STM implementations flush both log entries and data updates to PM, and we consider the time spent in flushing log entries as part of flushing overheads.
Clearly, flushing overheads are the biggest performance bottlenecks in these applications.
%In fact, flushing overheads are the biggest performance bottlenecks in these applications as logging overheads are comparatively much smaller.

\vspace{-4mm}
\begin{figure}[h]
    \centering
    \myincludegraphics{3.5in}{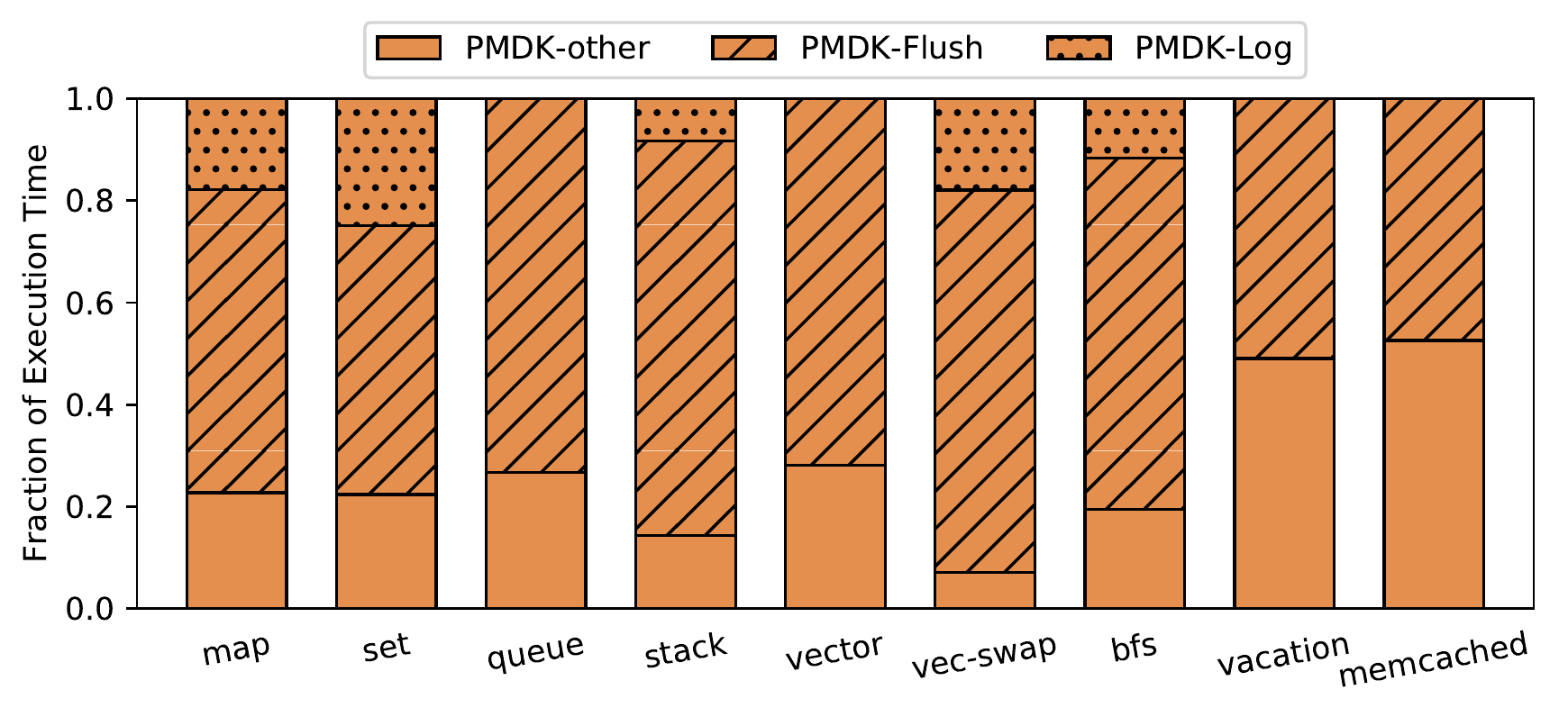}
    \caption{Fraction of execution time spent logging and flushing data in PM workloads using PMDK v1.5.}
    \label{fig:tx-overheads}
\end{figure}
\vspace{-4mm}

As we show in the rest of this section, the high flushing overheads in PM-STM are caused by excessive ordering constraints (\sfence) limit the overlapping of long-latency flush instructions.
Undo-logging techniques typically require 5-50 fences~\cite{nalli:2017} per transaction.
These fences mainly order log updates before the corresponding data updates.
In some implementations, the number of fences per transaction scales with the number of modified cachelines.
In our workloads with hybrid undo-redo logging, we observed 4-23 flushes and 5-11 fences per transaction (Figure~\ref{fig:flushes} in Section~\ref{sec:eval}).
Consequently, the median number of flushes overlapped per fence is 1-2, resulting in high flush overheads.

\begin{figure}[t]
    \centering
    \myincludegraphics{2.5in}{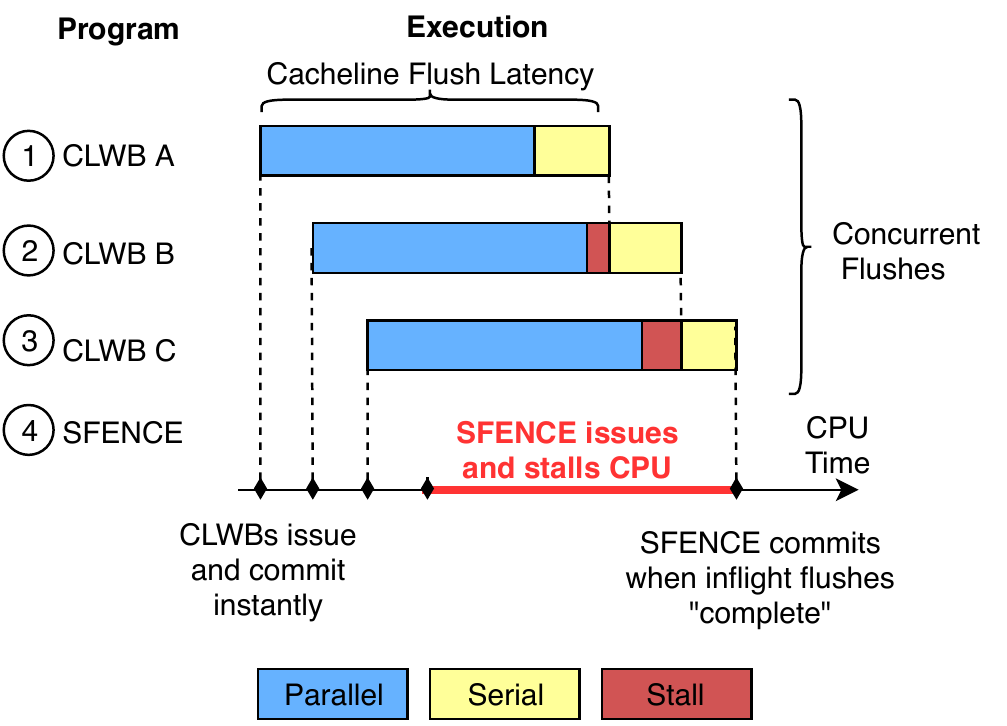}
    \caption{Execution of concurrent flushes on Optane DCPMM.}
    \label{fig:flush-timeline}
\end{figure}

\mypar{Flushes on Test Machine.}
\label{subsec:flush-intro}
In this paper, we focus on the \clwb instruction that writes back a dirty cacheline but may not evict it from the caches.
This instruction commits instantly but launches a cacheline flush that proceeds in the background of execution, unordered with other flushes to different addresses, as shown in Figure~\ref{fig:flush-timeline}.
Ordering points (\sfence) stall the CPU until all inflight flushes are completed.
On our test machine (described in Table~\ref{table-config}), we observed the latency of one \clwb followed by one \sfence to be 353 ns when the address being flushed was present in the L1D cache.
Thus, ordering points degrade performance by bringing the flush latency of weakly ordered flushes on the critical path.
In the rest of this paper, we use the term flushes to refer to weakly ordered flushes.

\begin{figure}[th]
    \centering
    \myincludegraphics{3in}{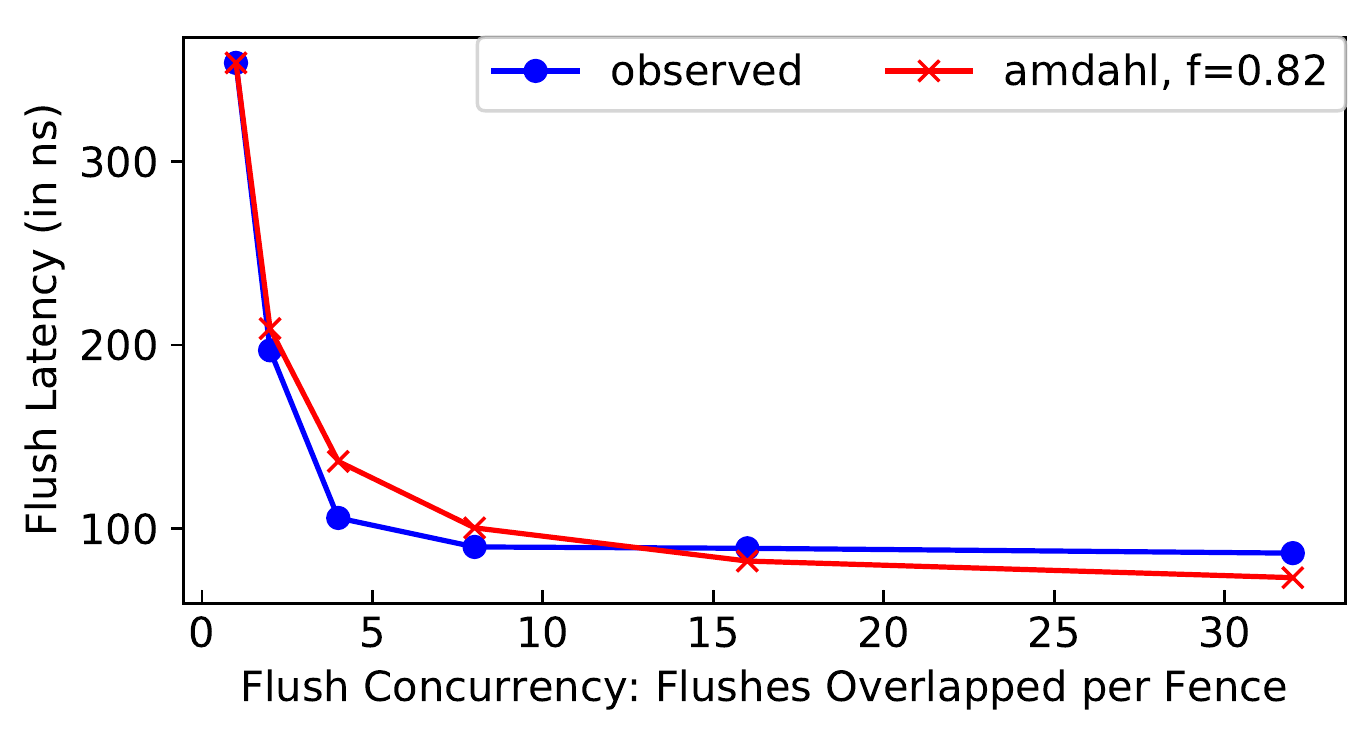}
    %\vspace{-4mm}
    \caption{Average Latency of PM cacheline flush observed on Optane DCPMM and estimated by analytical model (amdahl).}
    \label{fig:amdahl}
    %\vspace{-6mm}
\end{figure}

\mypar{Effects of Ordering Points.}
To mitigate the high flush overheads, we must reduce the frequency of ordering points and enable the overlap of multiple flushes.
We evaluated the efficacy of this approach on Optane DCPMMs via a simple microbenchmark.
Our microbenchmark first allocates an array backed by PM.
It issues writes to 320 random cachelines (= 20KB < 32 KB L1D cache) within the array to fault in physical pages and fetch these cachelines into the private L1D cache.
Next, it measures the time taken to issue \clwb instructions to each of these cachelines.
Fence instructions are performed at regular intervals e.g., one \sfence after every $N$ \clwb instructions.
The total time (for 320 \clwb + variable \sfence instructions) is divided by 320 to get the average latency of a single cacheline flush.
Figure~\ref{fig:amdahl} reports the average flush latency for varying flush concurrency.

The blue line in Figure~\ref{fig:amdahl} shows that the average flush latency can be effectively reduced by overlapping flushes, up to a limit.
Compared to a single un-overlapped flush (\clwb+\sfence), performing 16 flushes concurrently reduces average flush latency by 75\%.
However, performing 32 flushes concurrently only reduces average flush latency by 3\% compared to the case with 16 concurrent flushes.
Beyond 32, there is no noticeable improvement in flush latency.

\mypar{Analytical Model of Flush Latencies.}
As a side note, while Figure~\ref{fig:amdahl}’s blue line gives the empirical benefit of overlapping flushes, it also seems to closely follow Amdahl's law~\cite{Amdahl:1967}.
In particular, the red line shows an Amdahl’s law fit using the Karp-Flatt metric~\cite{Karp:1990} that has concurrent flushes acting 82\% parallel and 18\% serial.
With the 18\% serial component, it is easy to understand the diminishing returns of many concurrent flushes.
As the hardware is a black box, we do not yet know what features cause the appearance of serialization in the system under test.

\section{Minimally Ordered Durable Datastructures}
\label{body}
We address the high flushing costs with Minimally Ordered Durable (MOD) datastructures that allow failure-atomic and durable updates to be performed with \emph{one ordering point} in the common case.
These datastructures significantly reduce flushing overheads that are the main bottleneck in recoverable PM applications.
We have five goals for these datastructures:
\begin{enumerate}
    \item \emph{Failure-atomic updates} for recoverable applications. 
    \item \emph{Minimal ordering constraints} to tolerate flush latency.
    \item \emph{Simple programming interface} that hides implementation details for handling simple use cases.
    \item \emph{Allow composition} of failure-atomic updates to multiple datastructures.
    \item \emph{Support for common datastructures} for application programmers such as set, map, vector, queue and stack.
    \item \emph{No hardware modifications} needed to enable high performance applications on currently available systems.
\end{enumerate}

We first introduce the \emph{Functional Shadowing} technique underpinning MOD datastructures.
Next, we show a recipe to create MOD datastructures from existing functional datastructures.
Then, we describe MOD's programming interfaces.

\subsection{Functional Shadowing}
\begin{figure}[t]
    %\hspace{2cm}
    \centering
    \myincludegraphics{3.5in}{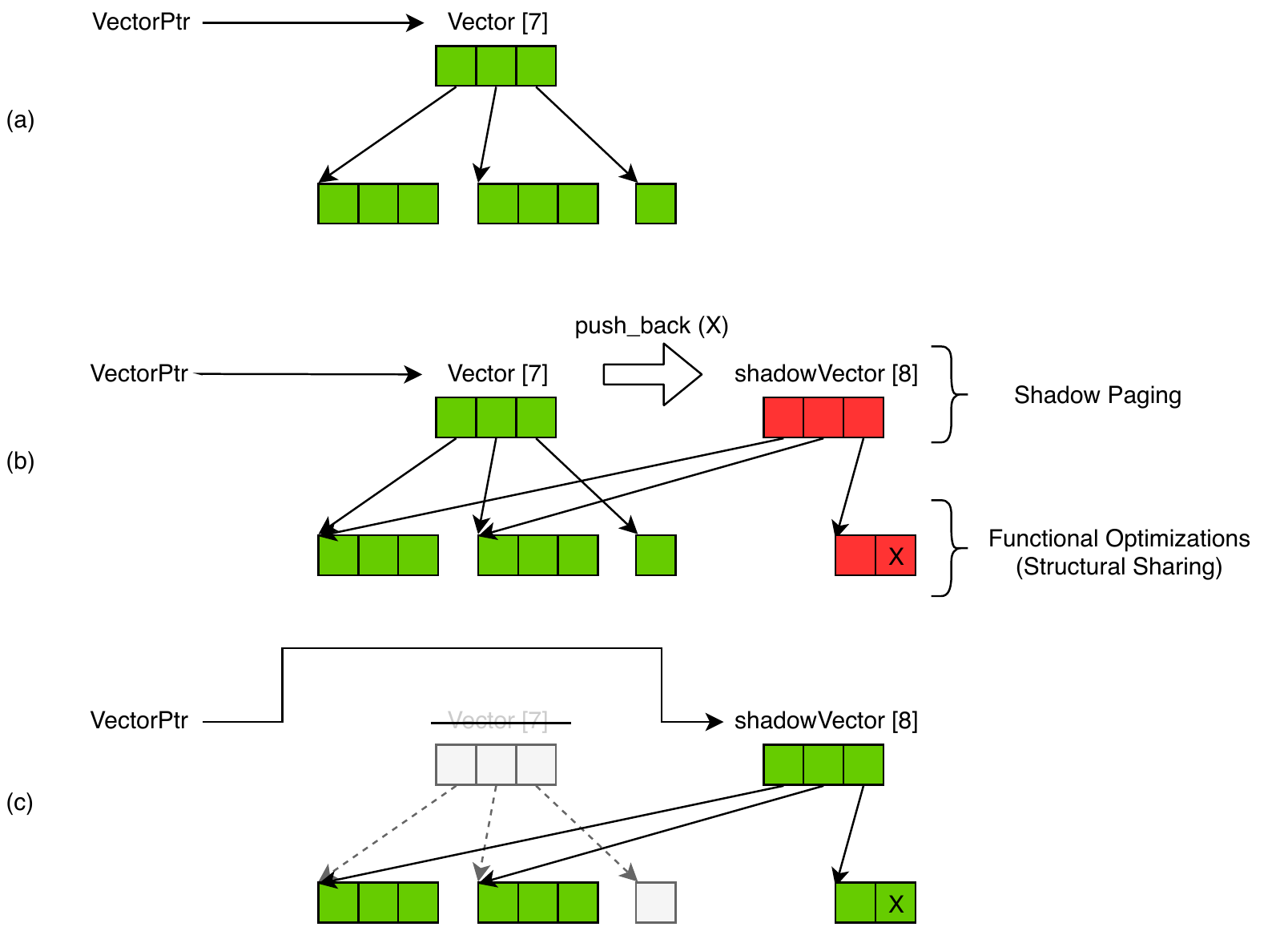}
    \caption{Functional Shadowing in action on (a) MOD vector. (b) Shadow is created on Append (i.e., \code{push\_back}) operation that reuses data from the original vector. (c) Application starts using updated shadow and old data is cleaned up.}
    \label{fig:fs}
\end{figure}

Functional Shadowing leverages shadow paging techniques to minimize ordering constraints in updates to PM datastructures and uses optimizations from functional datastructures to reduce the overheads of shadow paging.
As per shadow paging techniques, we implement non-destructive and out-of-place update operations for all MOD datastructures.
Accordingly, updates of MOD datastructures logically return a new version of the datastructure without any modifications to the original data.
As shown in Figure~\ref{fig:fs}, a \code{push\_back} operation in a vector of size 7 would result in a new version of size 8 while the original vector of size 7 remains untouched.
We refer to the updated version of the datastructure as a \emph{shadow} in accordance with conventional shadow paging techniques.

There are no ordering constraints in creating the updated shadow as it is not considered a necessary part of application state yet.
We do not log these writes as they do not overwrite any useful data.
In case of a crash at this point, recovery code can reclaim  memory corresponding to any partially updated shadow in PM.
Due to the absence of ordering constraints, we can overlap flushes to all dirty cachelines comprising the updated shadow to minimize flushing overheads.
A single ordering point is sufficient to ensure the completion of all the outstanding flushes and guarantee the durability of the shadow.
Subsequently, the application must atomically replace the original datastructure with the updated shadow.
For this purpose, we offer multiple efficient Commit functions described in the next subsection.
In contrast, PM-STM implementations perform in-place modifications which overwrite existing data and need logging to revert partial updates in case of crashes.
In-place updates also introduce ordering constraints as log writes must be ordered before the corresponding data update.

We reduce shadow paging overheads using optimizations commonly found in functional datastructures.
Conventional shadow paging techniques incur high overheads as the original data must be copied completely to create the shadow.
Instead, we use \emph{structural sharing} optimizations to maximize data reuse between the original datastructure and its shadow copy.
We illustrate this in Figure~\ref{fig:fs}, where \code{shadowVector} reuses 6/8 internal nodes from the original \code{Vector} and only adds 2 internal and 3 top-level nodes.
In the next subsection, we discuss a method to convert existing implementations of functional datastructures to MOD datastructures.

\subsection{Recipe for MOD Datastructures}

We provide a simple recipe for creating MOD datastructures out of existing implementations of functional datastructures:
\begin{enumerate}
    \item First, we use an off-the-shelf persistent memory allocator \code{nvm{\textunderscore}malloc}~\cite{nvm_malloc} to allocate datastructure state in PM.
    \item Next, we allocate the internal state of the datastructure on the persistent heap instead of the volatile stack.
    \item Finally, we extend all update operations to flush all modified PM cachelines with \clwb instructions and no ordering points. These flushes will be ordered by an ordering point in a Commit step described later in this section.
\end{enumerate}

While functional datastructures do not support durability by default, they offer a suitable starting point from which to generate MOD datastructures.
They support non-destructive update operations which are typically implemented through pure functions.
Thus, every update returns a new updated version (i.e., shadow) of the functional datastructure without modifying the original.
They export simple interfaces such as map, vector, etc. that are implemented internally as highly optimized trees such as Compressed Hash-Array Mapped Prefix-trees~\cite{Steindorfer:2015} (for map, set) or Relaxed Radix Balanced Trees~\cite{Stucki:2015} (for vector).
These implementations are designed to amortize the overheads of data copying as needed to create new versions on updates.

Optimized functional implementations also have low space overheads via structural sharing, i.e., maximizing data reuse between the original data and the shadow.
Tree-based implementations are particularly amenable to structural sharing.
On an update, the new version creates new nodes at the upper levels of the tree, but these nodes can point to (and thus reuse) large sub-trees of unmodified nodes from the original datastructure.
The number of new nodes created grows extremely slowly with the size of the datastructures, resulting in low overheads for large datastructures.
As we show in our evaluation section, the additional memory required on average for an updated shadow is less than 0.01\% of the memory of the original datastructure of size 1 million elements.

Moreover, the trees are broad but not deep to avoid the problem of `bubbling-up of writes'~\cite{Condit:2009} that plagues conventional shadow paging techniques.
This problem arises as the update of an internal node in the tree requires an update of its parent and so on all the way to the root.
We find that existing implementations of such functional datastructures are commonly available in several languages, including C++ and Java.

We conjecture that the ability to create MOD datastructures from existing functional datastructures is important for three reasons.
First, we benefit from significant research efforts towards lowering space overheads and improving performance of these datastructures~\cite{Driscoll:1989,Okasaki:1998, Puente:2017, Steindorfer, Stucki:2015}.
Secondly, programmers can easily create MOD implementations of additional datastructures beyond those in this paper by using our recipe to port other functional datastructures.
Finally, we forecast that this approach can help extend PM software beyond C and C++ to Python, JavaScript and Rust, which have implementations of functional datastructures.

\subsection{Programming Interface}
\label{subsec:fase}
To abstract away the details of Functional Shadowing from application programmers, we provide two alternative interfaces for MOD datastructures:

\begin{itemize}
    \item A \textbf{Basic} interface that abstracts away the internal versioning and is sufficient for simple use cases.
    \item A \textbf{Composition} interface that exposes multiple versions of datastructures to enable complex use cases, while still hiding the complexities of the implementation.
% FS internals and is recommended for expert programmers. 
\end{itemize}

\begin{figure}[h]
    \centering
    \myincludegraphics{\columnwidth}{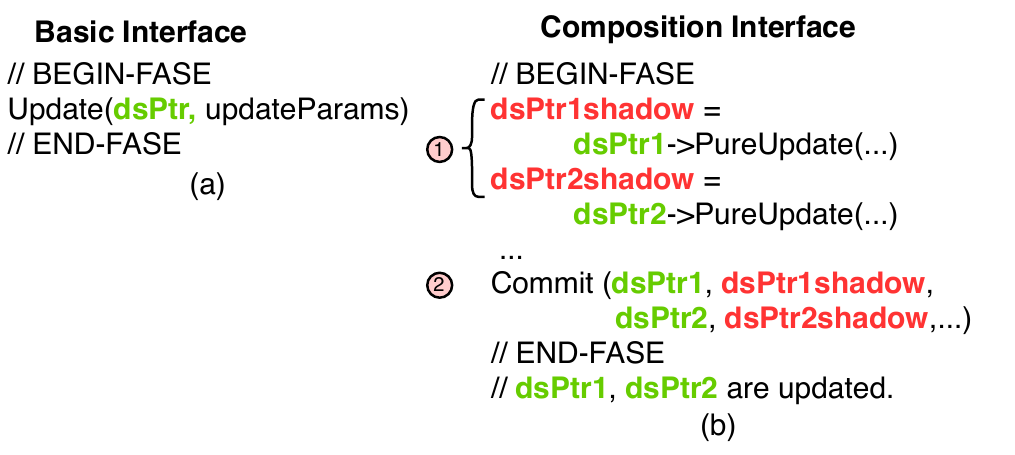}
    %\vspace{-8mm}
    \caption{Failure-atomic Code Sections (FASEs) with MOD datastructures using (a) Basic interface to update one datastructure and (b) Composition interface to atomically update multiple datastructures with (1) Update and (2) Commit steps.}
    \label{fig:interface}
    %\vspace{-4mm}
\end{figure}

\subsubsection{Basic Interface}
The Basic interface to MOD datastructures (Figure~\ref{fig:interface}a) allows programmers to perform individual failure-atomic update operations to a single datastructure.
With this interface, MOD datastructures appear as mutable datastructures with logically in-place updates.
Programmers use pointers to datastructures (e.g., \code{ds1} in Figure~\ref{fig:interface}a), as is common in PM programming.
Each update operation is implemented as a self-contained FASE with one ordering point, as described later in the next section.
If the update completes successfully, the datastructure pointer points to an updated and durable datastructure.
In case of crash before the update completes, the datastructure pointer points to the original durable and uncorrupted datastructure.
We expose common update operations for datastructures such as \code{push\_back}, \code{update} for vectors, \code{set} for sets/maps, \code{push}, \code{pop} for stacks and \code{enqueue}, \code{dequeue} for queues, as in C++ STL.

\begin{figure*}[h]
    \centering
    \myincludegraphics{6.5in}{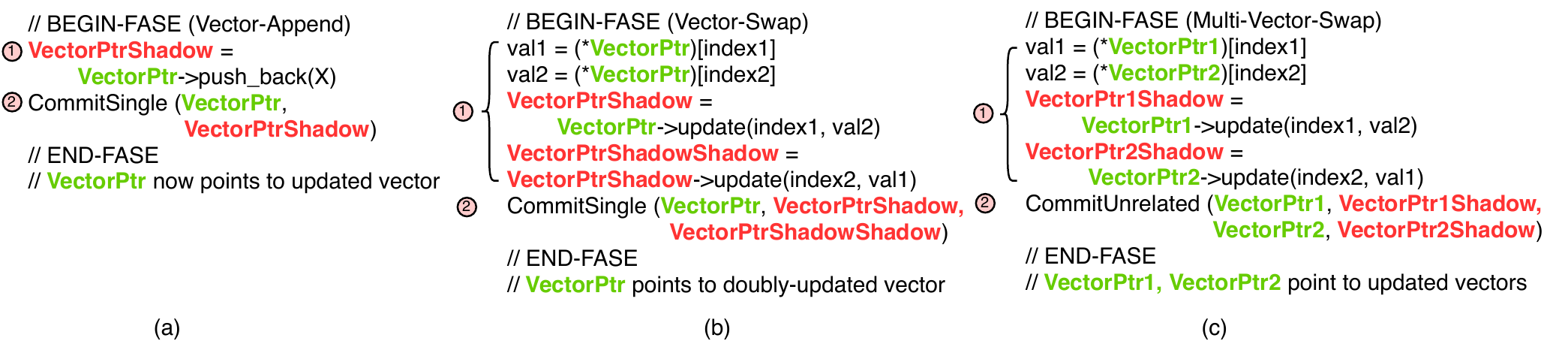}
    %\vspace{-6mm}
    \caption{Using the Composition interface for failure-atomically (a) appending an element to a vector, (a) swapping two elements of a vector and (c) swapping two elements of two different vectors.}
    \label{fig:interface-multi-example}
    %\vspace{-3mm}
\end{figure*}

The Basic interface targets the common case when a FASE contains only one update operation on one datastructure.
This common case applies to all our workloads except \code{vacation} and \code{vector-swaps}.
For instance, \code{memcached} relies on a single recoverable map to implement its cache and FASEs involve a single \code{set} operation.

\ignore{
The Basic interface can also handle more complex use cases if programmers , albeit with the restriction that all failure-atomic updates be performed as part of a single datastructure operation, i.e., function call.
For such purposes, programmers have to go  bespoke functions have to be added to the datastructure implementation for each particular use case.
For instance, we would have to add an operation to swap the value of two indices in a vector for the workload \code{vector-swaps}.
To minimize such changes to the underlying datastructure implementations, we advocate the use of Composition datastructures for complex use cases, described next.
}

\subsubsection{Composition Interface}
The Composition interface to MOD datastructures (Figure~\ref{fig:interface}b) is a general-purpose transaction-like programming interface.
It allows programmers to failure-atomically perform updates on multiple datastructures or perform multiple updates to the same datastructure or any combination thereof.
For instance, moving an element from one queue to another requires a pop operation on the first queue and a push operation on the second queue, both performed failure-atomically in one FASE.
Complex operations such as swapping two elements in a vector also require two update operations on the same vector to be performed failure-atomically.
In such cases, the Composition interface allows programmers to perform individual non-destructive update operations on multiple datastructures to get new versions, and then atomically replace all the updated datastructures with their updated versions in a single \emph{Commit} operation.

With this interface, programmers can build complex FASEs, each with multiple update operations on multiple datastructures.
Each FASE must consist of two parts: Update and Commit.
During Update, programmers perform updates on one or more MOD datastructures.
On an update operation, the original datastructure is preserved and a new updated version is returned that is guaranteed to be durable only after Commit.
Thus, programmers are temporarily exposed to multiple versions of datastructures.
Programmers use the Commit function to atomically replace all the original datastructures with their latest updated and durable versions.
Our Commit implementation (described in Section~\ref{subsec:fase-impl}) contains a single ordering point in the common case.
We use this interface in two workloads: \code{vector-swaps} and \code{vacation}. 

Figure~\ref{fig:interface-multi-example} demonstrates the following use cases:

\mypar{Single Update of Single Datastructure:}  
While this case is best handled by the Basic interface, we repeat it here to show how this can be achieved with the Composition interface.        
In Figure~\ref{fig:interface-multi-example}a, appending an element to \code{VectorPtr} results in an updated version (\code{VectorPtrShadow}).
The Commit step atomically modifies \code{VectorPtr} to point to \code{VectorPtrShadow}.
As a result of this FASE, a new element is failure-atomically appended to \code{VectorPtr}.
    
\mypar{Multiple Update of Single Datastructure:} 
We show a FASE that swaps two elements of a vector in Figure~\ref{fig:interface-multi-example}b.
The Update step involves two vector lookups and two vector updates.
        The first vector update results in a new version \code{VectorPtrShadow}.
        The second vector update is performed on the new version to get another version (\code{VectorPtrShadowShadow}) that reflects the effects of both updates.
        Finally, Commit makes \code{VectorPtr} point to the latest version.

\mypar{Single Updates of Multiple Datastructures:} 
Figure~\ref{fig:interface-multi-example}c shows how we swap elements from two different vectors in one FASE.
For each vector, we perform update operation to get a new version.
In Commit, both vector pointers are atomically updated to point to the respective new versions.

\mypar{Multiple Updates of Multiple Datastructures:} The general case is realized by combining the previous use cases.

\section{Implementation Details}
\label{sec:mod-impl}
We now discuss our implementation of MOD datastructures.

\subsection{Implementation of Programming Interfaces}
\label{subsec:fase-impl}
We prioritize minimal ordering constraints in our implementation of the two interfaces to MOD datastructures. 

\begin{figure*}[t]
    \centering
    \myincludegraphics{6.5in}{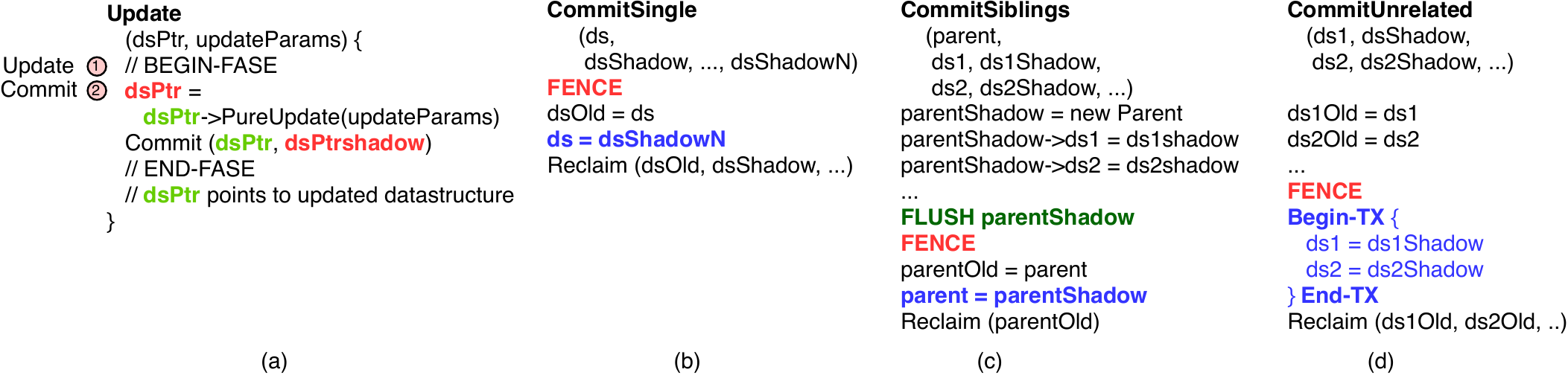}
    \caption{(a) Implementation of Basic interface as a wrapper around the Composition interface. Commit implementation shown for multi-update FASEs operating on (a) single datastructure, (b) multiple datastructures pointed to by common \emph{parent} object, and (c) (uncommon) multiple unrelated datastructures.}
    \label{fig:commit-impl}
    %\vspace{-5mm}
\end{figure*}

\mypar{Basic Interface.}
As shown in Figure~\ref{fig:commit-impl}a, the Basic interface is a wrapper around the Composition interface to create the illusion of a mutable datastructure.
The programmer accesses the MOD datastructure indirectly via a pointer.
On a failure-atomic update, we internally create an updated shadow of the datastructure by performing the non-destructive update.
Then, using Commit, we ensure the durability of the shadow and atomically update the datastructure pointer to point to the updated and durable shadow.
Thus, we hide FS details from the programmer.

\mypar{Composition Interface.}
The Composition interface can be used to build complex multi-update FASEs, each with one ordering point in the common case.

To support the Update step, MOD datastructure supports non-destructive update operations. 
Within these update operations, all modified cachelines are flushed using (weakly ordered) \clwb instructions and there are no ordering points or fences.
However, this step results in multiple versions of the updated MOD datastructures.

The Commit step ensures the durability of the updated versions and failure-atomically updates all relevant datastructure pointers to point to the latest version of each datastructure.
We provide optimized implementations of \code{Commit} for two common cases as well as the general implementation, as shown in Figure~\ref{fig:commit-impl}.
We discuss the memory reclamation needed to free up unused memory in Section~\ref{sec:mod-impl}.

The first common case (\code{CommitSingle} in Figure~\ref{fig:commit-impl}b) occurs when one datastructure is updated one or multiple times in a FASE (e.g., Figure~\ref{fig:interface-multi-example}a,b).
To Commit, we update the datastructure pointer to point to the latest version after all updates with a single 8B (i.e., size of a pointer) atomic write.
We then reclaim the memory of the old datastructure and intermediate shadow versions i.e., all but the latest shadow version.

The second common case (\code{CommitSiblings} in Figure~\ref{fig:commit-impl}c) occurs when the application updates two or more MOD datastructures that are pointed to by a common persistent object (\emph{parent}) in one FASE.
%For instance, the application may want to update multiple datastructures corresponding to different fields of the same \code{struct.
In this case, we create a new instance of the parent (\code{parentShadow}) that points to the updated shadows of the MOD datastructures.
Then, we use a single pointer write to replace the old parent itself with its updated version.
We used this approach in porting \code{vacation}, wherein a \code{manager} object has three separate recoverable maps as its member variables.
A commonly occurring parent object in PM applications is the root pointer, one for each persistent heap, that points to all recoverable datastructures in the heap.
Such root pointers allow PM applications to locate recoverable datastructures in persistent heaps across process lifetimes.

In these two common cases, our approach requires \textbf{only one ordering point per FASE}.
The single ordering point is required in the commit operation to guarantee the durability of the shadow before we replace the original data. 
The entire FASE is a single epoch per the epoch persistency model~\cite{Pelley:2014}.
Both of the common cases require an atomic write to a single pointer, which can be performed via an 8-byte atomic write.
In contrast, PM-STM implementations require 5-11 ordering points per FASE (Section~\ref{subsec:flush-conc}).

For the general and uncommon case (\code{CommitUnrelated} in Figure~\ref{fig:commit-impl}d) where two un-related datastructures get updated in the same FASE, we need to atomically update two or more pointers.
For this purpose, we use a very short transaction (STM) to atomically update the multiple pointers, albeit with more ordering constraints.
Even in this approach, the majority of the flushes are performed concurrently and efficiently as part of the non-destructive updates.
Only the flushes to update the persistent pointers in the Commit transaction cannot be overlapped due to PM-STM ordering constraints.

Thus, the Composition interface enables efficient FASEs that update multiple datastructures in the two common cases.

\subsection{Correctness}
\label{subsec:correct}
We provide a simple and intuitive argument for correct failure-atomicity of MOD datastructures.
The main correctness condition is that there must not be any pointer from persistent data to any unflushed or partially flushed data.
MOD datastructures support non-destructive updates that involve writes only to newly allocated data and so there is no possibility of any partial writes corrupting the datastructure.
All writes performed to the new version of the datastructure are flushed to PM for durability.
During Commit, one fence orders the pointer writes after all flushes are completed i.e., all updates are made durable.
Finally, the pointer writes in Commit are performed atomically.
If there is a crash before the atomic pointer writes in Commit, the persistent pointers point to the consistent and durable original version of the datastructure.
If the atomic pointer writes complete successfully, the persistent pointers points to the durable and consistent new version of the datastructures.
Thus, we support correct failure-atomic updates of MOD datastructures.

\subsection{Memory Reclamation}
Leaks of persistent memory cannot be fixed by restarting a program and thus are more harmful than leaks of volatile memory.
Such PM leaks can occur on crashes during the execution of a FASE.
Specifically, allocations from an incomplete FASE leak PM data that must be reclaimed by recovery code.
Additionally, our MOD datastructures must also reclaim the old version of the datastructure on completion of a successful FASE.

We use reference counting for memory reclamation.
Our MOD datastructures are implemented as trees.
In these trees, each internal node maintains a count of other nodes that point to it i.e., parent nodes.
We increment reference counts of nodes that are reused on an update operation and decrement counts for nodes whose parents are deleted on a delete operation.
Finally, we deallocate a node when its reference count hits 0.
%This process of memory reclamation may be performed on a background thread, similar to log-flushing optimizations in PM transactions~\cite{Kolli:2017}.

Our key optimization here is to recognize that reference counts do not need to be stored durably in PM.
On a crash, all reference counts in the latest version can be scanned and set to 1 as the recovered application sees only one consistent and durable version of each datastructure.

We rely on garbage collection during recovery to clean up allocated memory from an incomplete FASE (on a crash).
Our performance results include the time spent in garbage collection.
As our datastructures are implemented as trees, we can perform a reachability analysis starting from the root node of each MOD datastructure to mark all memory currently referenced by the application.
Any unmarked data remaining in the persistent heap is a PM leak and can be reclaimed at this point.
%There is a potential to extend functional shadowing to the PM allocator to prevent such leaks in the first place, but we leave this for future work.
A common solution for catching memory leaks is to log memory allocator activity.
%Thus on recovery, we can replay the log and free any memory allocated in the last FASE.
%However, this approach reintroduces some ordering and logging overheads.
However, this approach reintroduces ordering constraints and degrades the performance of all FASEs to prevent memory leaks in case of a rare crash.

\subsection{Automated Testing}
While it is tricky to test the correctness of recoverable datastructures, the relaxed ordering constraints of shadow updates allow us to build a simple and automated testing framework for our MOD datastructures.
We generate a trace of all PM allocations, writes, flushes, commits, and fences during program execution.
Subsequently, our testing script scans the trace to ensure that all PM writes (except those in commit) are only to newly allocated PM and that all PM writes are followed by a corresponding flush before the next fence.
By verifying these two invariants, we can test the correctness of recoverable applications as per our correctness argument in Section~\ref{subsec:correct}.
%One exception is reference counts, which we store in PM for convenience but which do not need to be durable (as explained in the previous section).
%We added debug prints in our library to trace addresses of updated reference counts.
%Our python tool discards these addresses from the valgrind-generated trace.

\ignore{
\subsection{Proposed Optimizations}
There are two general optimizations that can lead to further improvements in MOD datastructures that we have not implemented.

\mypar{Offline Memory Reclamation.}
If CPU resources are available, we can move memory reclamation off the critical path and onto a separate dedicated thread.
During \code{Commit}, the application adds a reference to the old datastructure to a per-thread reclamation queue instead of actively reclaiming the memory.
This lowers latency by eliminating the cost of decrementing reference counts and freeing memory from the main thread, but at the cost of throughput due to coordination overhead.
This optimization is similar to offline log truncation done in some PM-STM implementations~\cite{Huang:2014, Volos:2011}.

\mypar{In-place Shadow Updates.}
We can improve performance by performing some updates in-place without increasing ordering constraints.
When a MOD datastructure is modified more than once in a FASE, we can perform in-place updates for all modifications after the first one.
The first update creates a shadow copy that is not linked into the application state, and thus further modifications can be done in-place without any ordering or logging.
For example, in \code{vector-swaps} (Figure~\ref{fig:interface-multi-example}b), we can perform the second update in-place without incurring any shadow copying overheads.
}

\ignore{
\section{Discussion}
In this section, we seek to answer some questions readers may have about Functional Shadowing.

\mypar{Why make programmers learn functional programming?} \\
We hide the details of functional programming in the implementation of Functional Shadowing.
Programmers can directly use our library of recoverable datastructures without being exposed to its internal layout, similar to how programmers use the C++ Standard Template Library or Java Collections.
In fact, we modified \emph{mnemosyne} to directly use our recoverable map datastructure without incorporating Functional Shadowing at the application-level.

Note that we use pure functions solely to disallow in-place updates of persistent data and not for their inherent fault tolerance.
Pure functions are fault-tolerant as we can redo them with the same inputs and always get the same outputs.
However, it is difficult to ensure the validity and same-ness of the inputs after a crash. 
Others have successfully used the related concept of idempotent functions to implement PM transactions that can be rolled forward after a crash~\cite{ido}.

Functional Shadowing enables the rapid development of a software ecosystem for persistent applications by reusing research and engineering efforts spent on functional programming languages.
Specifically, it can be used to extend PM software development beyond C and C++ to Python, Javascript and Rust, which have support for pure functions and immutable datastructures.

Functional Shadowing can also enable new application domains such as persistent streaming analytics.
Streaming analytics allows real-time analytic processing on data streaming from different input sources such as social media or sensors.
Current streaming analytics frameworks use operators such as Map, Reduce, Filter using pure functions on immutable datastructures.
To allow recovery from system or OS failures, the state of worker nodes is checkpointed every few minutes.
The amount of data loss on a crash can be minimized with PM-backed streaming analytics with functional shadowing.

However, we must note that functional shadowing does have some limitations.
First, it does place additional stress on memory management and particularly memory allocation, which is known to be a challenge with persistent memory~\cite{Bhandari:2016, Oukid:2017}.
Second, applications with Functional Shadowing perform an increased number of stores to PM, compared to their transactional equivalents.
We measure the impact of these characteristics in Section~\ref{sec:eval}.
}
\ignore{
Second, while PM writes remain noticeable slower than DRAM writes, high-performance persistent datastructures can be built by reducing the state stored in PM~\cite{btree-papers}.
For instance, many persistent btrees only store leaf nodes in PM, while the indexes are stored in faster DRAM.
In case of crashes, recovery code regenerates the volatile indexes.
However, building such recoverable datastructures requires careful design choices for each individual datastructure.
Also, such optimizations are often orthogonal to the choice of crash-consistency mechanism used.
}

\section{Evaluation}
\label{sec:eval}
In this work, we seek to provide a library of recoverable datastructures with good abstractions and good performance.
We answer four questions in our evaluation:

\begin{enumerate}
    \setlength{\itemsep}{-.5pt}
    \item \textbf{Programmability}: What was our experience programming with MOD datastructures?
    \item \textbf{Performance}: Do MOD datastructures improve the performance of recoverable workloads compared to PM-STM?
    \item \textbf{Ordering Constraints}: Do workloads with MOD datastructures have fewer fences than with PM-STM?
    \item \textbf{Additional Overheads}: What are the additional overheads introduced by MOD datastructures?
\end{enumerate}

\subsection{Methodology}
\mypar{Test System Configuration.}
We ran our experiments on a machine with actual Persistent Memory---Intel Optane DCPMM~\cite{optane}---and upcoming second-generation Xeon Scalable processors (codenamed Cascade Lake).
We configured our test machine such that Optane DCPMM is in 100\% App Direct mode~\cite{intel-modes} and uses the default Directory protocol.
In this mode, software has direct byte-addressable access to the Optane DCPMM.
Table~\ref{table-config} reports relevant details of our test machine.
We measured read latencies using Intel Memory Latency Checker v3.6~\cite{intel-mlc}.

\begin{table}[h]
\centering
\resizebox{0.8\columnwidth}{!}{%
\begin{tabular}{@{}ll@{}}
\toprule
\multicolumn{2}{c}{\textbf{CPU}}                                \\ \midrule
    Type    &  Intel Cascade Lake  \\
    Cores                             & 96 cores across 2 sockets                         \\
    Frequency                         & 3.7 GHz (with Turbo Boost)                      \\
    Caches                            & \begin{tabular}[c]{@{}l@{}} L1: 32KB Icache, 32KB Dcache  \\ L2: 1MB, L2: 33 MB (shared) \end{tabular}                                                                \\ \midrule
    \multicolumn{2}{c}{\textbf{Memory System}}                      \\ \midrule
    PM Capacity & 2.9 TB (256 GB/DIMM) \\ 
    PM Read Latency & 302 ns (Random 8-byte read) \\
    DRAM Capacity & 376 GB \\
    DRAM Read Latency & 80 ns (Random 8-byte read) \\  \bottomrule
\end{tabular}%
    }
    \caption{Test Machine Configuration.}
    \label{table-config}
\end{table}

\mypar{Hardware Primitives.}
The Cascade Lake processors on our test machine support the new \clwb instruction for flushing cachelines.
The \clwb instruction flushes a dirty cacheline by writing back its data but may not evict it.
Our workloads use \clwb instructions for flushing cachelines and the \sfence instructions to order flushes.
%System configuration of the test machine is reported in Table~\ref{tbl-config}.

\mypar{OS interface to PM.}
Our test machine runs Linux v4.15.6.
The DCPMMs are exposed to user-space applications via the DAX-filesystem interface~\cite{lwn-dax}. 
Accordingly, we created an ext4-dax filesystem on each PM DIMM.
Our PM allocators create files in these filesystems to back persistent heaps.  
We map these PM-resident files into application memory with flags \codesm{MAP\_SHARED\_VALIDATE} and \codesm{MAP\_SYNC}~\cite{lwn-sync} to allow direct user-space access to PM.

\mypar{PM-STM Implementation.}
We use the PM-STM implementation (\emph{libpmemobj}) from Intel's PMDK library~\cite{pmdk} in our evaluations.
We choose PMDK as it is publicly available, regularly updated, Intel-supported and optimized for Intel's PM hardware.
Moreover, PMDK (v1.4 or earlier) has been used for comparison by most earlier PM proposals~\cite{Correia:2018, Liu:2017, ido, Shin:2017:PFF, Shin:2017}.
We evaluate both PMDK v1.5 (released October 2018), which uses hybrid undo-redo logging techniques as well as PMDK v1.4, which primarily relies on undo-logging.

\mypar{Workloads.}
Our workloads include several microbenchmarks and two recoverable applications, consistent with recent PM works~\cite{Correia:2018, Kolli:2017, Liu:2017, ido, Shin:2017:PFF, Shin:2017}.
As described in Table~\ref{benchmarks}, our microbenchmarks involve operations on commonly used datastructures: map, set, queue, list and vector.
The \code{vector-swaps} workload emulates the main computation in the \code{canneal} benchmark from the PARSEC suite~\cite{Bienia:2008}. 
The baseline map datastructure can be implemented by either hashmap or ctree from the WHISPER suite~\cite{nalli:2017}.
Here, we compare against hashmap which outperformed ctree on Optane DCPMM.
Moreover, we also measured two recoverable applications from the WHISPER suite: \code{memcached} and \code{vacation}.
We modified these applications to use the PMDK and MOD map implementations.
The only other PM-STM application in WHISPER is \code{redis}, but it also uses a map datastructure so we found it redundant for our purposes.
The other WHISPER benchmarks are \emph{not applicable} for our evaluation as they are either filesystem-based or do not use PM-STM. 
Instead, we created the \code{bfs} workload that uses a recoverable queue for breadth-first search on the large Flickr graph~\cite{uf-coll}. We do not store the graph durably but reconstruct it from the dataset in each execution.
We ran all workloads to completion on real hardware.

\begin{table}[]
    \centering
    \large
    \resizebox{1\columnwidth}{!}{
        \begin{tabular}{p{.7in}p{2.5in}p{1.4in}}
\hline
Benchmark    & Description                                                          & Configuration                                  \\ \hline
map   & Insert/Lookup random keys in map                        & 8B key, 32B value      \\
set   & Insert/Lookup random keys in set                                            & 8B key, 32B value   \\
stack   & Push/Pop elements from top of stack                              & 8B elements \\
queue   & Enqueue/Dequeue elements in queue & 8B elements                         \\
\hline
vector & Update/Read random indices in vector                          & 8B element  \\
vec-swap & Swap two random elements in vector            & 8B element \\
\hline
bfs & Breadth-First Search using recoverable queue on Flickr graph~\cite{uf-coll} & 0.82M nodes, 9.84M edges, 8B elements \\
vacation     & Travel reservation system with four recoverable maps                 & query range:80\%, 55\% user queries \\

    memcached    & In-memory key value store using one recoverable map                  & 95\% sets, 5\% gets, 16B key, 512B value                           \\ \hline
\end{tabular}
}
\caption{Benchmarks used in this study. Workloads performs 1 million iterations of the operations described.}
\label{benchmarks}
\end{table}

\subsection{Programmability}
While the rest of this section presents quantitative performance results, in this paragraph we qualitatively describe the programmability of MOD datastructures.
We demonstrated the use of MOD datastructures in two existing applications: \code{vacation} and \code{memcached}.
With MOD datastructures as with C++ STL, applications get access to datastructures via narrow, expressive interfaces but without access to the internal implementation.
However, \code{memcached}, like many PM applications, uses a custom datastructure (hashmap) whose implementation is tightly coupled to the application logic.
Thus our main challenge was to decouple the code (i.e., application logic) from the internal datastructure implementation.
We do not expect this to be an issue when building new applications.
For instance, \code{vacation} was easy to port as its datastructure implementations were neatly encapsulated.
Also, \code{vacation}'s logic required composing failure-atomic updates to multiple distinct maps that were members of the same object, for which we used our Composition interface with \code{CommitSiblings}. 

\subsection{Performance}

\begin{figure}[t]
    \centering
    \myincludegraphics{3.5in}{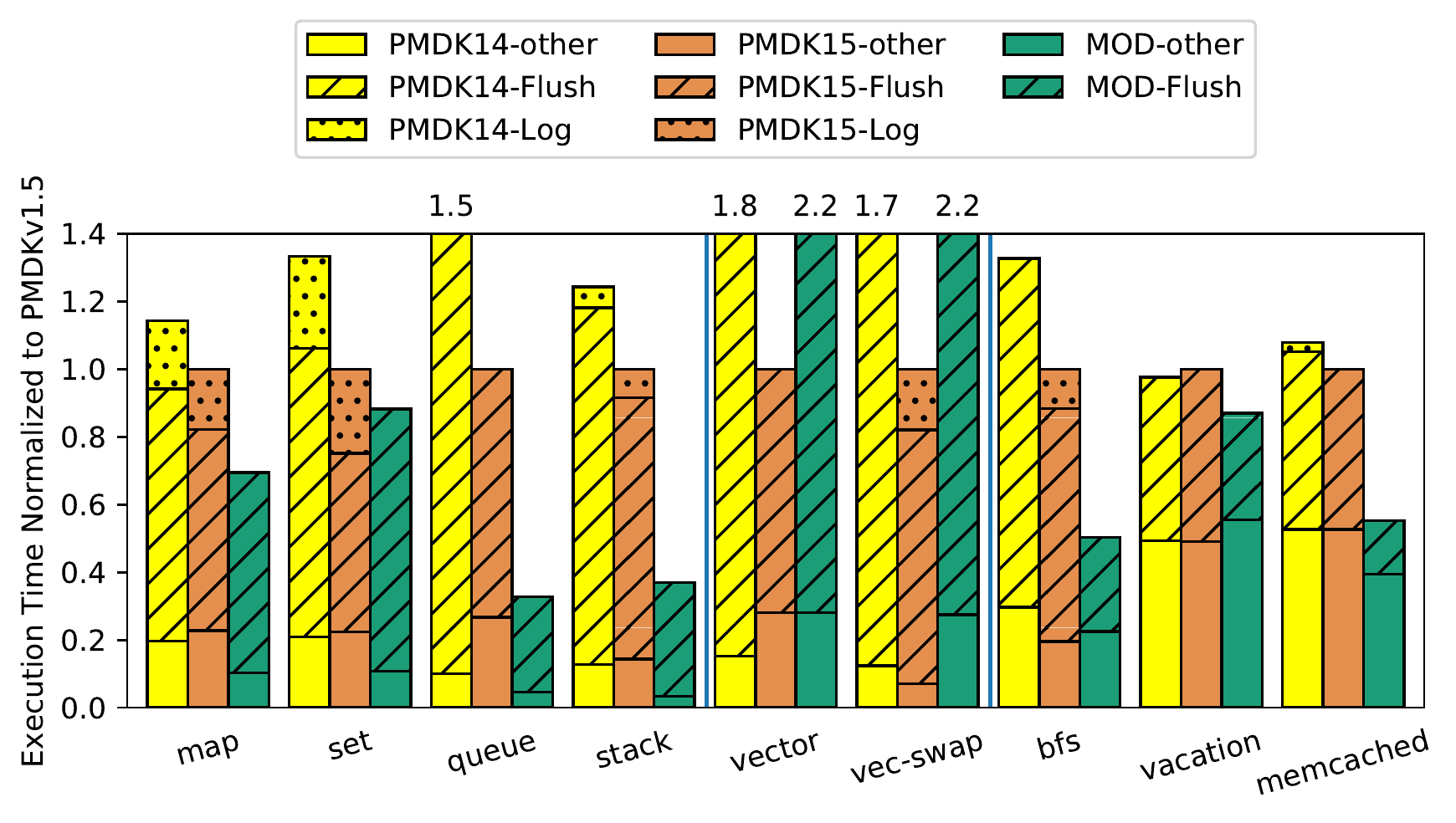}
    %\vspace{-5mm}
    \caption{Execution Time of PM workloads, normalized to PMDK v1.4 implementation of each workload.}
    \label{fig:perf}
    %\vspace{-5mm}
\end{figure}

Figure~\ref{fig:perf} shows the execution time (so smaller is better) of PM workloads with PMDK transactions and MOD datastructures. 
We make the following observations.

First, PMDK v1.5 with hybrid undo-redo logging performs 23\% better on average than undo-logging based PMDK v1.4, due to optimizations targeting transaction overheads~\cite{pmdk15}.

Second, MOD datastructures offer a speedup of 43\% on average for pointer-based datastructures (map, set, queue, stack) over PMDK v1.5. The performance improvements are attributed to lower flushing overheads (50\% vs 66\% of PMDK v1.5 execution time) and no logging overheads (0\% vs 13\%).

Third, for only vector and \code{vec-swap} microbenchmarks, the abstraction benefits of MOD datastructures come with a performance cost—not benefit.
This occurs due to the overhead of moving from a dense 1-D array to a tree-based implementation that functional datastructures use to facilitate incremental updates. 
Future work can examine whether this slowdown can be mitigated or that the MOD abstractions must have a performance cost in this case.

For the three full applications, MOD datastructures show an average speedup of 36\% over PMDK v1.5.
Here, the performance improvements arise from lower flushing overheads (25\% vs 50\% of PMDK v1.5 execution time) and no logging overheads.
\code{vacation} shows a lower speedup of 13\% as we have to copy and flush the parent object in our approach (with \code{CommitSiblings}), while PMDK can update in place.

\subsection{Flushing Concurrency}
\label{subsec:flush-conc}
\begin{figure}[t]
    \centering
    \myincludegraphics{3in}{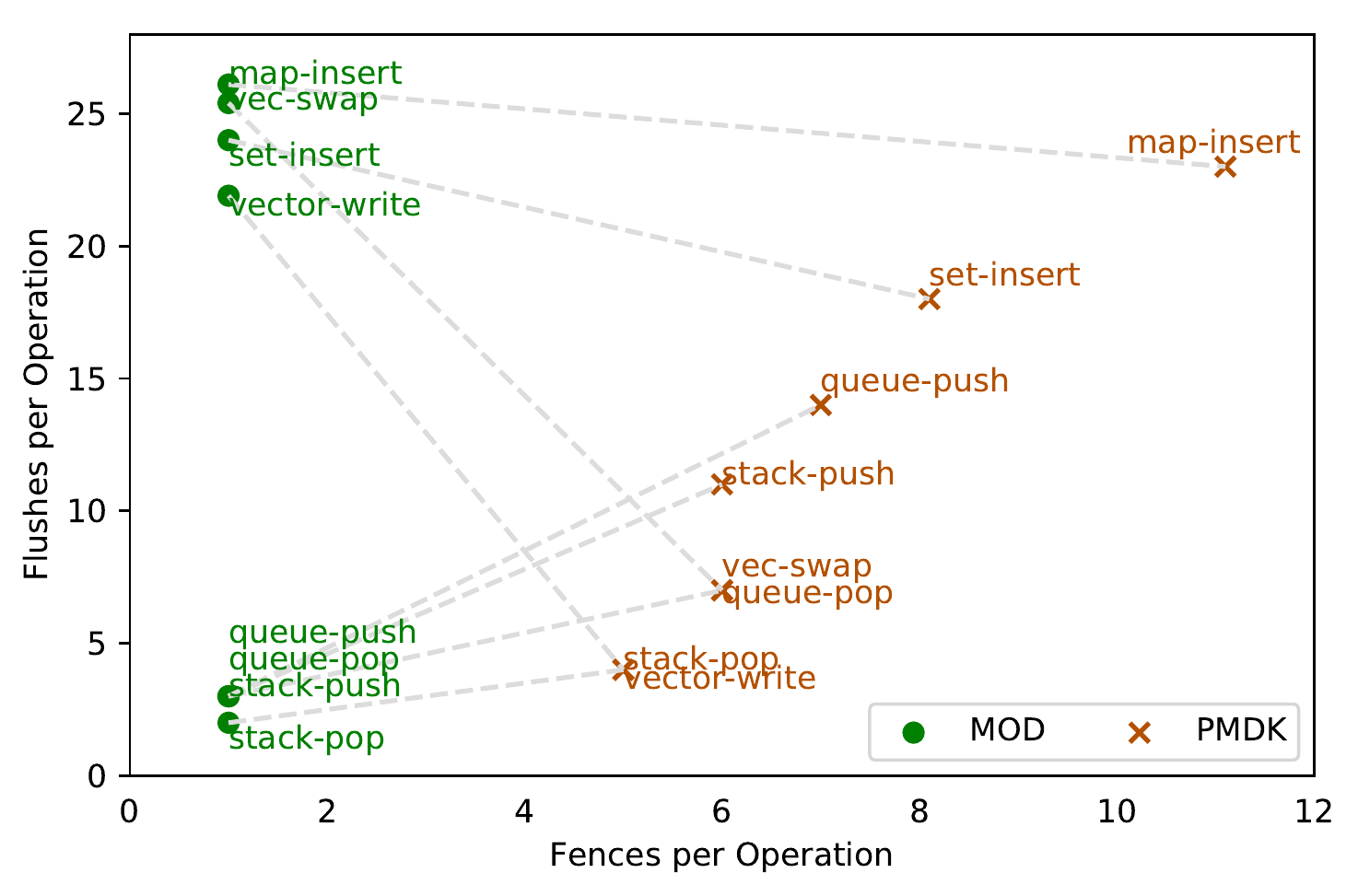}
    \caption{Flush and fence frequency in PM workloads.}
    \label{fig:flushes}
\end{figure}

Figure~\ref{fig:flushes} illustrates ordering (x-axis) and flushing frequency (y-axis) in update operations to evaluated datastructures.
Lookup operations in our workloads do not require any flush or fence instructions.
Here, we restrict the comparison to PMDK v1.5 and MOD.

Lower ordering constraints enable lower flushing overheads, if the number of flushed cachelines is comparable.
PMDK workloads typically exhibit a high number of fences per operation.
In our evaluated workloads, MOD datastructures always have only one fence per operation.
While MOD datastructures copy and flush additional data than PMDK, there are no log entries to be flushed.
For map and set implementations, the amount of flushed data is comparable in both approaches.
Pop operations in the MOD queue occasionally require a reversal of one of the internal linked lists resulting in greater flushing activity than PMDK on average.
The reduction in flushes comes from the absence of log entries as well as implementation differences.
However, writes and swaps to the MOD vector require significantly more cachelines to be flushed as compared to the PMDK vector.
This helps explain the performance degradation explained in the previous subsection.

\subsection{Additional Overheads}
MOD datastructures introduce two new overheads: space overheads and increased cache pressure.
First, extra memory is allocated on every update for the shadow, resulting in additional space overheads.
Secondly, functional datastructures (including vectors and arrays) are implemented as pointer-based datastructures with little spatial locality.

\mypar{Space overheads.}
In Table~\ref{tbl:memory}, we report the increase in memory consumption on doubling the capacity of datastructures, i.e., inserting an additional 1 million elements in a datastructure of size 1 million elements.
On average for most of our workloads (except vector), the memory consumption of MOD datastructures only grows 21\% faster than PMDK datastrucures.
More importantly, every individual update operation only requires 0.00002-0.00004$\times$ extra memory beyond the original version, as compared to 2$\times$ extra memory in naive shadow paging.
Thus, structural sharing in our datastructure implementations minimizes the FS space overheads.

\vspace{-3mm}
\begin{table}[h]
    \tiny
    \centering
    \resizebox{\columnwidth}{!}{%
    \begin{tabular}{|c|c|c|c|c|c|}
        \hline
        & map     & set    & stack   & queue  & vector\\ \hline
        MOD & 1.87$\times$ & 2.08$\times$ & 2.25$\times$ & 1.67$\times$ & 131$\times$ \\ \hline
        PMDK & 1.78$\times$ & 1.75$\times$ & 1.50$\times$ & 1.50$\times$ & 2$\times$ \\ \hline
    \end{tabular}
    }
    \caption{Ratio of memory consumed by datastructure with 2M elements compared to 1M elements.}% For PMDK map, set and vector, update operations happen in place without allocating memory.}
\label{tbl:memory}
\end{table}
\vspace{-5mm}

\mypar{Cache Pressure.} 
While our MOD datastructures typically perform better than PMDK datastructures, interestingly they also exhibit greater cache misses.
Unfortunately, it was not possible to separate cache misses to PM and DRAM in our experiments on real hardware, but we expect most of the cache accesses to be for PM cachelines in our workloads. 

\begin{figure}[th]
    \centering
    \myincludegraphics{3.5in}{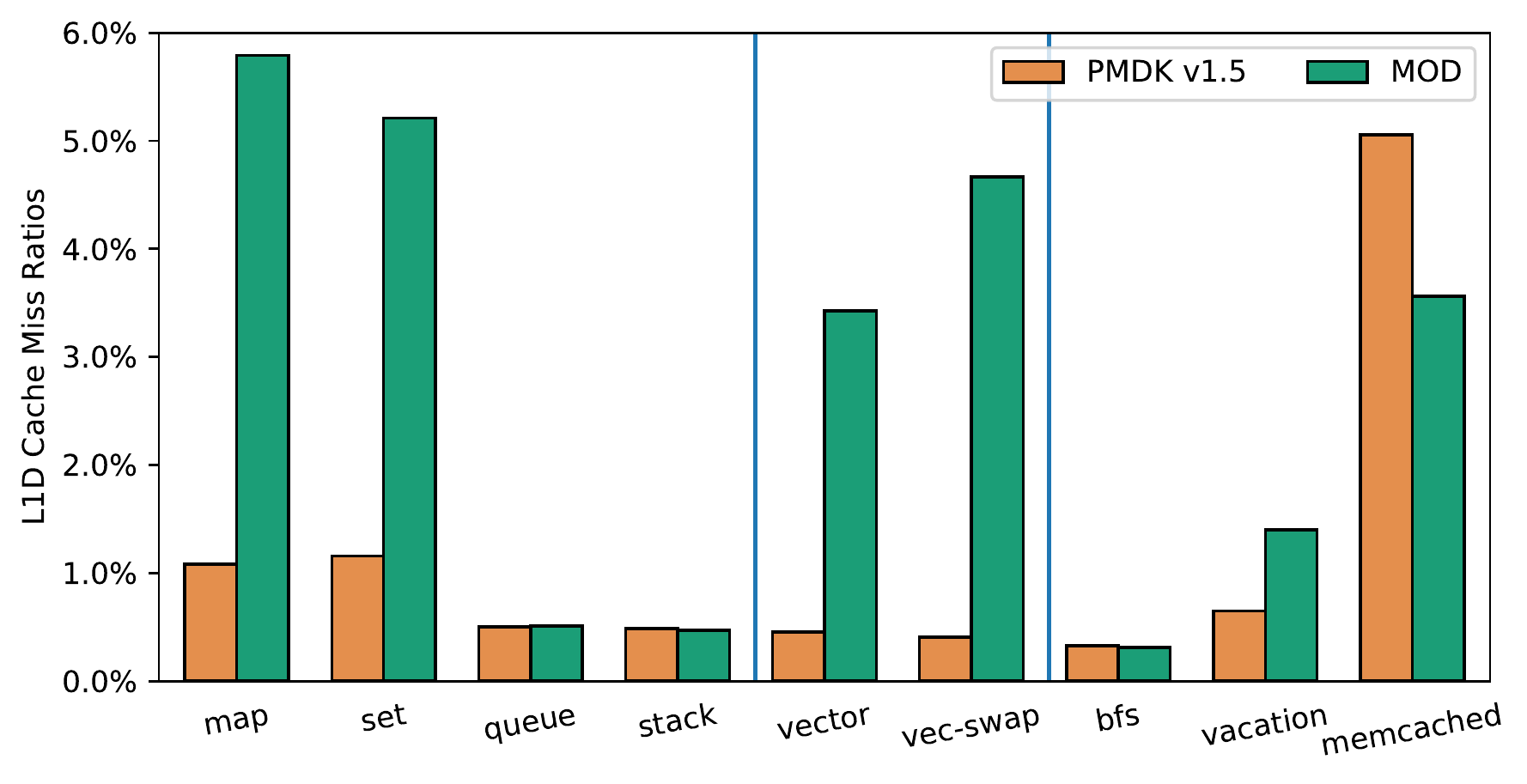}
    %\vspace{-8mm}
    \caption{L1D Cache miss ratios for PM workloads.}
    \label{fig:misses}
\end{figure}

The pointer-based functional implementations results in more cache misses particularly in the small L1D cache, as seen in Figure~\ref{fig:misses}.
This is evident in case of map, set and vector workloads, which show 2.8-4.6$\times$ the cache misses with MOD datastructures than with PMDK.
The PMDK implementations of map, set and vector involve arrays contiguously laid out in memory and thus have greater spatial locality and fewer pointer-chasing patterns.
However, the pointer-based implementations of MOD datastructures are necessary to reduce the shadow copying overheads.

MOD implementations of stack, queue and \code{bfs} show low cache miss ratios, comparable to the PMDK implementations.
These results to be expected as stacks and queues are pointer-based datastructures in both PMDK and MOD implementations.
Moreover, push and pop operations in these datastructures only operate on the head or the tail, resulting in high temporal locality of accesses.

\section{Related Work}
Prior works mainly consists of general optimizations for PM-STM and specific optimizations for durable datastructures.

\subsection{PM-STM Optimizations.}
Software approaches include Mnemosyne~\cite{Volos:2011}, NV-Heaps~\cite{Coburn:2011}, SoftWrap~\cite{SoftWrap}, Intel PMDK~\cite{pmdk}, JUSTDO~\cite{Izraelevitz:2016}, iDO~\cite{ido}, Romulus~\cite{Correia:2018}, DudeTM~\cite{Liu:2017}.
Mnemosyne, SoftWrap, Romulus and DudeTM rely on redo logging, NV-Heaps employs undo logging techniques and PMDK recently switched from undo logging (in v1.4) to hybrid undo-redo log (in v1.5)~\cite{pmdk-log}.
Each of these approaches requires \textbf{4+ ordering points per FASE}.
Most undo-logging implementations require ordering points proportional to the number of contiguous data ranges modified in each transaction and can have as many as 50 ordering points in a transaction~\cite{nalli:2017}.
In contrast, redo-logging implementations require relatively constant number of ordering points regardless of the size of the transaction and are better for large transactions.
However, redo logging requires load interposition to redirect loads to updated PM addresses, resulting in slow reads and increased complexity.
%As we have shown in this work, the high frequency of ordering points reduces the potential for minimizing overheads by overlapping flushes.
%Functional Shadowing requires only one ordering point per FASE.

Romulus and DudeTM both utilize innovative approaches based on redo-logging and shadow paging to reduce ordering constraints.
Romulus uses a volatile redo-log with shadow data stored in PM while DudeTM uses a persistent redo-log with shadow data stored in DRAM.
Both of these approaches double the memory consumption of the application as two copies of the data are maintained.
This is a greater challenge with DudeTM as the shadow occupies DRAM capacity, which is expected to be much smaller than available PM.
Our MOD datastructures only have two versions during an update operation, with significant data reuse between the two versions.
Both DudeTM and Romulus incur logging overheads and require store interposition, unlike MOD datastructures.

The optimal ordering constraints for PM-STM implementations under idealized scenarios have been analyzed~\cite{Kolli:2016}.
The results show that PM-STM performance can be improved using new hardware primitives that support Epoch or Strand Persistency~\cite{Kolli:2016}, neither of which are currently supported by any architectures.
In contrast, MOD datastructures reduce ordering constraints on currently available hardware.

Finally, better hardware primitives for ordering and durability have also been proposed.
For instance, DPO~\cite{DPO} and HOPS~\cite{nalli:2017} propose lightweight ordering fences that do not stall the CPU pipeline.
Efficient Persist Barriers~\cite{Joshi:2015} move cacheline flushes out of the critical path of execution by minimizing epoch conflicts.
Speculative Persist Barriers~\cite{Shin:2017} allow the core to speculatively execute instructions in the shadow of an ordering point.
Forced Write-Back~\cite{Steal} proposes cache modifications to perform efficient flushes with low software overheads.
All these proposals reduce the performance impact of each ordering point in PM applications, whereas we reduce the number of ordering points in these applications.
Moreover, these proposals require hardware modifications to the core and/or the cache hierarchy while MOD datastructures improve performance on unmodified hardware.

\subsection{Recoverable Datastructures.}
While Functional Shadowing provides a way to directly convert existing functional datastructures into recoverable ones, the following papers demonstrate the value of handcrafting recoverable datastructures.
Dali~\cite{nawab:2017} is a recoverable prepend-only hashmap that is updated non-destructively while preserving the old version.
Updates in both Functional Shadowing and Dali are logically performed as a single epoch to minimize ordering constraints.
However, our datastructures are optimized to reuse data between versions, while the Dali hashmap uses a list of historical records for each key.
The CDDS B-tree~\cite{Venkataraman:2011} is a recoverable datastructure that also relies on versioning at a node-granularity for crash-consistency.
%Version numbers are stored in the nodes of the B-tree to allow nondestructive updates but additional work is done on writes to order nodes by version number for faster lookup.
However, it is not straightforward to extend such fine-grained versioning to other datastructures beyond B-trees.
Instead, we rely on versioning at the datastructure-level.

There have also been several attempts at optimizing recoverable B+-trees, which are commonly used in key-value stores and filesystems.
NV-Tree~\cite{Yang:2015} achieves significant performance improvement by storing internal tree nodes in volatile DRAM and reconstructing them on a crash.
wB+-Trees~\cite{Chen:2015} uses atomic writes and bitmap-based layout to reduce the number of PM writes and flushes for higher performance.
These optimizations cannot be directly extended to other datastructures such as vectors and queues.
Our MOD datastructures are all implemented as trees, and could allow these optimizations to apply generally to more datastructures with further research.

\section{Conclusion}
Persistent memory devices are close to becoming commercially available.
Ensuring consistency and durability across failures introduces new requirements on programmers and new demands on hardware to efficiently move data from volatile caches into persistent memory.
Minimally ordered durable datastructures provide an efficient mechanism that leverages the performance characteristics of Intel’s Optane DCPMM for much higher performance.
Rather than focusing on minimizing the amount of data written, MOD datastructures minimize the ordering points that impose long program delays.
Furthermore, they can be created via simple extensions to a large library of existing highly optimized functional datastructures providing flexibility to programmers.

\bibliographystyle{plain}
\bibliography{main}

\end{document}